\documentclass[11pt]{article}

\usepackage{amsmath}
\usepackage{amssymb}
\usepackage[paper=a4paper,text={17cm,24cm},centering]{geometry}
\usepackage{graphicx}
\usepackage{rotating}
\usepackage{hyperref}
\hypersetup{colorlinks=true, citecolor=red, urlcolor=blue, linkcolor=blue}

\def\be{\begin{equation}}
\def\ee{\end{equation}}

\def\BKS{BK${}_{\mathfrak S}$~}
\def\BKT{BK${}_{\mathfrak T}$~}

\begin{document}

\title{Analytical asymptotics for hard diffraction}
\author{{Anh Dung Le${}^{(1)}$, Alfred H. Mueller${}^{(2)}$,
  St\'ephane Munier${}^{(1)}$}\\
  {\footnotesize\it  (1) CPHT, CNRS, \'Ecole polytechnique, IP Paris,
    F-91128 Palaiseau, France}\\
  {\footnotesize\it  (2) Department of Physics, Columbia University,
  New York, NY 10027, USA}}
\date{March 19, 2021}

\maketitle

\begin{abstract}
 We show that the cross section for diffractive dissociation of a small onium off a large nucleus at total rapidity $Y$ and requiring a minimum rapidity gap $Y_{\text{gap}}$ can be identified, in a well-defined parametric limit, with a simple classical observable on the stochastic process representing the evolution of the state of the onium, as its rapidity increases, in the form of color dipole branchings: It formally coincides with twice the probability that an even number of these dipoles effectively participate in the scattering, when viewed in a frame in which the onium is evolved to the rapidity $Y-Y_{\text{gap}}$. Consequently, finding asymptotic solutions to the Kovchegov-Levin equation, which rules the $Y$-dependence of the diffractive cross section, boils down to solving a probabilistic problem. Such a formulation authorizes the derivation of a parameter-free analytical expression for the gap distribution. Interestingly enough, events in which many dipoles interact simultaneously play an important role, since the distribution of the number $k$ of dipoles participating in the interaction turns out to be proportional to $1/[k(k-1)]$.
\end{abstract}

%%%%%%%%%%%%%%%%%%%%%%%%%%%%%%

\setcounter{tocdepth}{2}

\tableofcontents

%%%%%%%%%%%%%%%%%%%%%%%%%%%%%%

\section{Introduction}

Diffraction has been observed in the scattering of
protons and nuclei~\cite{Alberi:1981af,Goulianos:1982vk};
more unexpectedly, also in deep-inelastic
electron-proton
scattering~\cite{Ahmed:1995ns,Derrick:1995wv,Schoeffel:2009aa}.
Diffractive events
should represent a sizable fraction of the events seen
at future electron-ion colliders~\cite{Accardi:2012qut,Agostini:2020fmq},
on the order of 20-30\%~\cite{Bendova:2020hkp}.

At high energies, electron-hadron scattering cross sections
may always be calculated starting from onium-hadron cross sections.
Indeed, in an appropriate frame,
the electron-hadron interaction is mediated by a colorless quark-antiquark
(onium)
fluctuation of a virtual photon picked in the state
of the electron~\cite{Nikolaev:1990ja,Ewerz:2006vd}
(see e.g. Ref.~\cite{Kovchegov:2012mbw} for an overview of high energy QCD).

Diffractive events are traditionally split into two classes:
Quasi-elastic scattering events, in which the
diffractive system typically consists in a vector meson or in a hadronized
open quark-antiquark pair,
and high-mass diffractive
dissociation events, in which the diffractive system possesses an invariant
mass on the order of the
center-of-mass energy of the onium-hadron subreaction, and a sizable
multiplicity.
While the former have drawn a lot of attention recently
(see e.g.~Ref.~\cite{Mantysaari:2020axf} for a review),
less effort has been devoted to the latter.
In many works, the high-mass diffractive system is treated as a quark-antiquark-gluon
system~(see e.g.~\cite{GolecBiernat:1998js,Bendova:2020hkp}),
neglecting any further quantum evolution. While this is a fair approximation
for phenomenological studies, the description of the parametric
large-rapidity asymptotics requires to account for all possible
fluctuations.

An equation for the distribution of the rapidity gap in the form
of an evolution in the total rapidity~$Y$, here refered to as the Kovchegov-Levin
(KL) equation, was established
in Ref.~\cite{Kovchegov:1999ji},
but only numerical solutions have been known~\cite{Levin:2001yv}.
Recently, based on a simple partonic picture of diffraction in which a gap
of size $Y_\text{gap}$ is due to an unusual fluctuation in the onium
evolution at rapidity $Y-Y_\text{gap}$,
the asymptotic functional form of that distribution
was argued~\cite{Mueller:2018zwx,Mueller:2018ned}, in the
limit of large rapidities and in the geometric scaling~\cite{Stasto:2000er}
kinematical region (see also Ref.~\cite{Contreras:2018adl} for
an alternative calculation).
But the overall numerical factor was believed not to be computable.
In the present paper, we shall provide a parameter-free expression for the gap
distribution, based on a
new formulation of diffractive dissociation in terms of a probabilistic
process.

Our starting point (Sec.~\ref{sec:formulation})
will be the well-known relation between fixed impact-parameter
semi-inclusive cross sections and
$S$-matrix elements solutions of the
Balitsky-Kovchegov (BK) equation~\cite{Balitsky:1995ub,Kovchegov:1999ua}.
In an appropriate limit, we shall derive
a probabilistic formulation of diffraction, involving
weights of the number of color-singlet effective independent exchanges
between the diffractive system and the nucleus.
Analytical expressions will be obtained for the latter~(Sec.~\ref{sec:analytics}),
from which the diffractive cross section
and the rapidity gap distribution will be derived
(Sec.~\ref{sec:diff}). We present
our conclusions and some prospects in Sec.~\ref{sec:conclusion}, and we report on
on a numerical check of our results in the appendix.

%%%%%%%%%%%%%%%%%%%%%%%%%%%%%%%%%%%%%%%%%%%%%%%%%%%%%%%%%%%%%%%%%%

\section{\label{sec:formulation}Formulation of scattering cross sections}

We consider the scattering of an onium of initial size $r$ off a large nucleus. The relative
rapidity of these colliding objects is denoted by $Y$.
In the following, we shall rescale all the rapidity variables by multiplying them
by $\bar\alpha\equiv \alpha_s N_c/\pi$. The kinematical rapidities will be
written in capital letters, and the rescaled rapidities in lowercase letters;
For example, we shall replace the total relative rapidity by $y\equiv \bar\alpha Y$.

Throughout this paper, we shall rely on the color dipole picture~\cite{Mueller:1993rr}
to represent the Fock state of an onium of fixed initial size $r$
by a random set of dipoles of various sizes $\{r_i\}$.
It assumes the limit of infinite number $N_c$ of colors, and is relevant
in the large-rapidity limit. Technically, the dipole model can
be seen as a tool to resum systematically
all planar graphs that contribute to the probability of a given Fock state,
keeping the terms that
dominate when the onium is viewed in a frame in which it is very fast.

The distribution of these sets of dipoles depends on the rapidity of the onium in the
reference frame
in which it is probed. In the dipole picture, Fock states
are elegantly generated by a simple $1\rightarrow 2$
branching process in rapidity~\cite{Mueller:1993rr}, which
is a particular branching random walk (see e.g. Ref.~\cite{Munier:2009pc} for a review).
The latter is defined by the rate of branching per unit rapidity 
$dp_{1\rightarrow 2}(\underline{r},\underline{r}')$ of
a dipole of size vector $\underline{r}$ into a pair of dipoles of size vectors
$\{\underline{r}',\underline{r}-\underline{r}'\}$
(the common endpoint of these dipoles being localized within
a surface of size $d^2\underline{r}'$), which reads in QCD
\be
dp_{1\rightarrow 2}(\underline{r},\underline{r}')\equiv \frac{d^2\underline{r}'}{2\pi}
\frac{r^2}{{r'}^2|\underline{r}-\underline{r}'|^2}.
\ee

\subsection{Matrix element for onium-nucleus forward elastic scattering}

We shall express the onium-nucleus cross sections with the help of the forward elastic
scattering matrix element ${\cal S}$
for a given set of dipoles present in the state of the onium
at rapidity $\tilde y_0\equiv y-y_0$. At high energies, cross sections are purely
absorptive, hence, in the conventions we shall use, the scattering amplitudes and
the $S$-matrix elements are all real.
Given that the dipoles are assumed to interact independently with the nucleus,
we can write
\be
   {\cal S}(y_0)=\prod_{\{r_i\}}S(y_0,r_i),
   \label{eq:product}
\ee
where $S(y_0,r_i)$ is the $S$-matrix element that encodes the
scattering of a single dipole
of size $r_i$ off the nucleus (averaged over the fluctuations of the target
nucleus), at relative rapidity $y_0$ (which is
the rapidity of the nucleus in the chosen frame.)
This matrix element solves the BK evolution equation
in rapidity~$y_0$~\cite{Kovchegov:1999ji},
with as an initial condition at $y_0=0$ the onium-nucleus $S$-matrix element
at zero relative rapidity taken e.g. from the McLerran-Venugopalan
model~\cite{McLerran:1993ni}.

We recall that the BK equation is a
non-linear integro-differential
equation which reads, for a function ${\mathfrak{S}}(y,{{r}})$,
\be
\partial_y{\mathfrak S}(y,{r})
=\int_{\underline{r}'}dp_{1\rightarrow 2}(\underline{r},\underline{r}')\left[
  {\mathfrak{S}}(y,{{r}'}){\mathfrak{S}}(y,|{\underline{r}-\underline{r}'}|)
  -{\mathfrak{S}}(y,{{r}})
  \right].
\label{eq:BKS}
\ee
The initial condition is given in the form of a function of $r$ only.
We will refer to Eq.~(\ref{eq:BKS}) as ``\BKS''.
The function $S(y_0,r_i)$ that appears in the right-hand side of Eq.~(\ref{eq:product})
solves \BKS, with the identification ${\mathfrak{S}}\equiv S$.
One may also write equivalently
the BK equation for the function $\mathfrak{T}\equiv 1-\mathfrak{S}$:
\be
\partial_y{\mathfrak T}(y,{r})
=\int_{\underline{r}'}dp_{1\rightarrow 2}(\underline{r},\underline{r}')\left[
  {\mathfrak{T}}(y,{{r}'})+{\mathfrak{T}}(y,|{\underline{r}-\underline{r}'}|)
  -{\mathfrak{T}}(y,{{r}})
  -{\mathfrak{T}}(y,{{r}'}){\mathfrak{T}}(y,|{\underline{r}-\underline{r}'}|)
  \right].
\label{eq:BKT}
\ee
Any equation in this form will be refered to as ``\BKT''.
Its solution is not known analytically. However, it has been established
that for a wide class of initial conditions, which includes all the ones
of interest for us, it tends asymptotically to a traveling wave~\cite{Munier:2003vc}.
Let us recall the main properties of this asymptotic solution.

We first need to introduce some background and a few useful notations.
The integral kernel of the linearized equation~(\ref{eq:BKT})
admits power functions of the form ${\mathfrak T}_\gamma(y,r)\equiv r^{2\gamma}$
as eigenfunctions. The eigenvalue equation reads
\be
\int_{\underline{r}'}dp_{1\rightarrow 2}(\underline{r},\underline{r}')\left[
  {\mathfrak{T}}_\gamma(y,{{r}'})+{\mathfrak{T}}_\gamma(y,|{\underline{r}-\underline{r}'}|)
  -{\mathfrak{T}}_\gamma(y,{{r}})\right]=\chi(\gamma)\,{\mathfrak T}_\gamma(y,{r}),
\ee
where
$\chi(\gamma)\equiv 2\psi(1)-\psi(\gamma)-\psi(1-\gamma)$, with the definition
$\psi(\gamma)\equiv d\ln\Gamma(\gamma)/d\gamma$.
Of particular importance for the physics we are investigating
are the eigenvalues in the vicinity of the peculiar one $\chi(\gamma_0)$,
where $\gamma_0\in]0,1[$ solves $\chi'(\gamma_0)=\chi(\gamma_0)/\gamma_0$
    \cite{Gribov:1984tu,Mueller:2002zm}.

It will prove convenient to use, instead of the dipole size variable~$r$,
a logarithm of $r$. More precisely, we define
\be
x\equiv\ln\frac{1}{r^2 Q_A^2},
\ee
and call any such quantity a ``log inverse size''.
In this definition, $Q_A$ is a fixed momentum scale
characteristic of the nucleus. (It can be identified with the
saturation momentum of the nucleus at rest, which emerges for example
from the McLerran-Venugopalan model~\cite{McLerran:1993ni}).

The traveling wave that solves the \BKT equation at large rapidities
is a smooth function connecting the fixed points
$1$ at $x\rightarrow -\infty$ and $0$ at
$x\rightarrow +\infty$. For a wide class of ``steep-enough'' initial conditions,\footnote{%
  ${\mathfrak T}(y=0,x)$ has to decrease faster than $e^{-\gamma_0 x}$
  when $x\rightarrow\infty$. Such solutions determined by the small-${\mathfrak T}$ tail
  are called ``pulled fronts'' in the
  terminology of Ref.~\cite{VANSAARLOOS200329}.
  }
to which all the initial conditions we will need to consider belong, many
properties of the traveling wave are independent of the latter.
The transition between~${\mathfrak T}=1$ and~${\mathfrak T}=0$ is located around the value
\be
X_y\equiv \chi'(\gamma_0)y-\frac{3}{2\gamma_0}\ln y
\ee
of $x$, within a region of typical size $1/\gamma_0$ (up to a non-universal
$y$-independent term, and up to terms vanishing for large $y$).
$X_y$ is related to the rapidity-dependent saturation momentum $Q_s(y)$ through
\be
X_y\equiv \ln\frac{Q_s^2(y)}{Q_A^2}.
\label{eq:Xy}
\ee
The shape of ${\mathfrak T}$ ahead of this transition region reads
\be
{\mathfrak T}(y,x)\simeq \text{const}\times (x-X_y)\,e^{-\gamma_0(x-X_y)}
\exp\left(-\frac{(x-X_y)^2}{2\chi''(\gamma_0)y}
\right),
\label{eq:BK-solution}
\ee
which holds in the limit of large $y$, and for values of $x$
such that $1\ll x-X_y\lesssim \sqrt{y}$.

Let us introduce the number density $n(x')$ of dipoles
of log inverse size~$x'$ in a given realization of the onium Fock state evolved to
the rapidity $\tilde y_0$. In terms of~$n$, ${\cal S}(y_0)$ in Eq.~(\ref{eq:product})
becomes
\be
{\cal S}(y_0)=
\prod_{x'}\left[S(y_0,x')\right]^{n(x') dx'},
\label{eq:S-as-product}
\ee
where the product now goes over all the bins in dipole log inverse size of
infinitesimal width~$dx'$. This also reads
\be
{\cal S}(y_0)=
\exp\left(-\int {dx'}\,n(x')\ln \frac{1}{S(y_0,x')}
\right)
\equiv
e^{-I({y_0})},
\ee
with the definition
\be
I(y_0)\equiv \int {dx'}\,{n(x')}\,{\ln \frac{1}{S(y_0,x')}}.
\label{eq:I-def}
\ee
Note that $I(y_0)$ is a random number, since $n$ is a random distribution.

The $S$-matrix element for the scattering of the initial onium
of log inverse size~$x$ at total relative rapidity $y$ reads
\be
S(y,x)=\left\langle
{\cal S}(y_0)\right\rangle_{\tilde y_0,x},
\label{eq:S-S}
\ee
where the averaging is over all the dipole configurations
of the onium at rapidity $\tilde y_0$,
namely over all realizations of $n$.
The obtained function $S$ is the same as the ones entering
the expression of ${\cal S}$ in Eq.~(\ref{eq:S-as-product}),
evaluated at a different rapidity.

%%%%%%

\subsection{Observables}

Let us recall the relation between cross sections for onium-nucleus scattering
at a fixed impact parameter $\underline{b}$
per unit transverse surface $d^2 \underline{b}$,
and $S$-matrix elements.
\begin{itemize}
  \item The total onium-nucleus cross section reads
\be
\sigma_{\text{tot}}(y,x)=2\left\langle 1-{\cal S}(y_0)\right\rangle_{\tilde y_0,x}.
\label{eq:sigmatot-exact}
\ee
 (In this formula and in all the following ones, the $\underline{b}$-dependence
in ${\cal S}$ and in the $\sigma$'s is understood).
The total cross section is obviously independent of the rapidity $y_0$,
although it is not manifest in the right-hand side of this equation.

\item The diffractive cross section, with a rapidity gap 
  not less than $y_0$, coincides with the elastic cross section
  for the scattering of the Fock state of the onium at rapidity $\tilde y_0$
  off the nucleus:
\be
\sigma_{\text{diff}}(y,x;y_0)=\left\langle\left[1-{\cal S}(y_0)\right]^2
\right\rangle_{\tilde y_0,x}.
\label{eq:sigmadiff-exact}
\ee
This cross section is the sum of the purely elastic and of the diffractive
dissociative cross sections.

\item The inelastic cross section is the difference between the total
cross section and the diffractive one. In terms of ${\cal S}$, it reads
\be
\sigma_{\text{in}}(y,x;y_0)=\left\langle 1-\left[{\cal S}(y_0)\right]^2
\right\rangle_{\tilde y_0,x}.
\label{eq:sigmain-exact}
\ee
\end{itemize}
At variance with $\sigma_{\text{tot}}$, the diffractive and inelastic cross sections 
obviously depend on $y_0$. Their rate of variation is related to the gap distribution, that we
define as
\be
\pi(y,r;y_\text{gap})\equiv
-\frac{1}{\sigma_\text{tot}}\left.\frac{\partial\sigma_\text{diff}}
{\partial y_0}\right|_{y_0=y_\text{gap}}
=\frac{1}{\sigma_\text{tot}}\left.\frac{\partial\sigma_\text{in}}
{\partial y_0}\right|_{y_0=y_\text{gap}}.
\label{eq:def-pi}
\ee

So far, these expressions are fully accurate
when so-called ``fan diagrams'' dominate the calculation of the
$S$-matrix. This is the case in the dipole model for
QCD evolution of the onium Fock state,
and when the scattering is off a very large
nucleus.

%%%%%%%%%%%%%%%%%%%%%%%%%%%%%%%%%%%%%%%%%%%%%%%%%%%%%%%%%%%%%%

\subsection{Probabilistic picture}

We now assume that in all the Fock state realizations which effectively
contribute to cross sections, the scattering probability of each of the
individual dipole is very small. In particular, the probability that the
same dipole scatters more than once is negligible. 
Then $S(y_0,x')$ can be assumed to be close to $1$ for all relevant values
of $x'$ in the expression of $I$ defined in Eq.~(\ref{eq:I-def}). 
For this reason, writing the latter in terms of $T=1-S$,
we may keep the term linear in $T$, and drop higher powers of $T$.
We denote by $I^{(1)}(y_0)$ the resulting overlap:
\be
I^{(1)}(y_0)= \int {dx'}\,{n(x')}\,{T(y_0,x')}.
\label{eq:I-approx}
\ee
This integral corresponds to the sum of all possible 
diagrams in which one single dipole in one given realization of
the onium Fock state,
the content of which is fully encoded in the number density $n$,
interacts by coupling to a single color-singlet
gluon pair which mediates the interaction with the evolved nucleus.
We shall call the approximation leading to Eq.~(\ref{eq:I-approx})
the ``single-exchange approximation''.

We further define
\be
F_N({\cal I})=\frac{{\cal I}^N}{N!}\quad\text{and}\quad
G_k({\cal I})=F_k({\cal I})\,e^{-{\cal I}}.
\label{eq:FN_Gk}
\ee
$F_N[I^{(1)}(y_0)]$ is the quantum-mechanical amplitude corresponding to
the sum of all the diagrams in which $N$ dipoles present in the
Fock state exchange color singlets with the nucleus at relative rapidity $y_0$.
$G_k[I^{(1)}(y_0)]$ is $F_k[I^{(1)}(y_0)]$ endowed with
an extra $e^{-I^{(1)}(y_0)}$ factor that unitarizes it ($\sum_k G_k=1$), turning
it into a quantity that may be interpreted as a probability:
It represents the probability that when choosing scattering configurations with a weight
given by their amplitude $F_N[I^{(1)}(y_0)]$, one picks those in which exactly $k$
dipoles interact. (Note that if the exponential is expanded,
it is seen to resum an infinity of graphs).
$G_k[I^{(1)}(y_0)]$ is also the {probability} that the set of dipoles that eventually interact
with the nucleus, when the event is viewed
from the restframe of the latter, are offspring of exactly $k$ dipoles, at the rapidity $y_0$
relative to the nucleus. 
Both $F_N$ and $G_k$ are understood to be evaluated
{\it for a given Fock state realization of the onium}.

We can now reformulate the scattering observables defined in
Eqs.~(\ref{eq:sigmatot-exact}),(\ref{eq:sigmadiff-exact}),(\ref{eq:sigmain-exact})
in terms of these
probabilities averaged over the realizations of the onium Fock state:
\begin{itemize}
\item The total cross section~(\ref{eq:sigmatot-exact}) simply reads
\be
\sigma_{\text{tot}}(y,x)=2\left(1-\left\langle G_0[I^{(1)}(y_0)]
\right\rangle_{\tilde y_0,x}\right)=
2\sum_{k=1}^\infty w_k(y,x;y_0),
\ee
where the terms in the right-hand side are the average weights, defined as
\be
w_k(y,x;y_0)\equiv\left\langle G_k[I^{(1)}(y_0)]\right\rangle_{\tilde y_0,x}.
\label{eq:def-w_k}
\ee
Note that each weight $w_k$ individually
may a priori be frame-dependent, i.e. depend upon $y_0$,
although the total cross section is boost invariant.

\item The inelastic cross section~(\ref{eq:sigmain-exact})
can also be expressed in terms of $w_k$.
We write
\be
\sigma_\text{in}(y,x;y_0)=\left\langle \left(e^{I^{(1)}(y_0)}
-e^{-I^{(1)}(y_0)}\right)e^{-I^{(1)}(y_0)}\right\rangle_{\tilde y_0,x}
\ee
and expand the difference of the exponentials to get
\be
\sigma_\text{in}(y,x;y_0)=
2\sum_{k\ \text{odd}} w_k(y,x;y_0).
\label{eq:sigmain}
\ee

\item Likewise, the diffractive cross section~(\ref{eq:sigmadiff-exact})
is then just twice the weight of the graphs in which
an even number of dipoles interact:
\be
\sigma_\text{diff}(y,x;y_0)=
2\sum_{k\ \text{even}} w_k(y,x;y_0),
\label{eq:sigmadiff}
\ee
where it is understood that the term $k=0$ is excluded from this sum.
\end{itemize}

%%%%%%%%%%%%%%%%%%%%

\subsection{Comparison to the Kovchegov-Levin equation}

We are going to show that the expression of the inelastic cross section in
Eq.~(\ref{eq:sigmain}) in terms of the weights $w_k$
is consistent with the solution to the KL equation.

%%%

\subsubsection{Brief review of the Kovchegov-Levin equation}

In this paragraph, we shall release the single-exchange approximation to
review the exact equation for diffraction in the framework of the dipole
model.\\

The KL equation, established in Ref.~\cite{Kovchegov:1999ji},
can be expressed as an evolution equation (with the rapidity $y$)
of the physical
probability that there be no inelastic scattering between the
state of the onium at $\tilde y_0$ and the nucleus,
\be S_{\text{in}}(y,r;y_0)\equiv 1-\sigma_\text{in}(y,r;y_0)=
\langle \left[{\cal S}(y_0)\right]^2\rangle_{\tilde y_0,r}
\ee
(see Eq.~(\ref{eq:sigmain-exact})).
The KL equation coincides with the \BKS (Eq.~(\ref{eq:BKS})) equation,
with the substitution ${\mathfrak S}=S_{\text{in}}$,
and the initial condition
\be
S_{\text{in}}(y_0,r;y_0)=\left[S(y_0,r)\right]^2.
\ee
Written in terms of the inelastic cross section $\sigma_\text{in}$ and
of the dipole amplitude $T$, this initial condition reads
\be
\sigma_{\text{in}}(y_0,r;y_0)=2T(y_0,r)-[T(y_0,r)]^2.
\label{eq:sigmain-IC-KL}
\ee
The rapidity gap distribution can easily be deduced from $S_\text{in}$, applying
Eq.~(\ref{eq:def-pi}) to $\sigma_{\text{in}}$.

%%%

\subsubsection{Evolution of the inelastic cross section
in the single-exchange approximation}

We now turn to the inelastic cross section as given
by Eq.~(\ref{eq:sigmain}), namely in the single-exchange approximation leading to
Eq.~(\ref{eq:I-approx}) for the overlap.

Let us define $W(y,r;y_0)$ to be
the weight of the graphs contributing to the onium-nucleus
forward elastic scattering amplitude
restricted to the diagrams in which there is an odd number
of dipoles interacting with the nucleus boosted to the rapidity $y_0$:
\be
W(y,r;y_0)\equiv \sum_{k\ \text{odd}} w_k(y,r;y_0).
\ee
The inelastic cross section~(\ref{eq:sigmain}) is just twice this quantity.

\begin{figure}
  \begin{center}
    \includegraphics[width=0.9\textwidth]{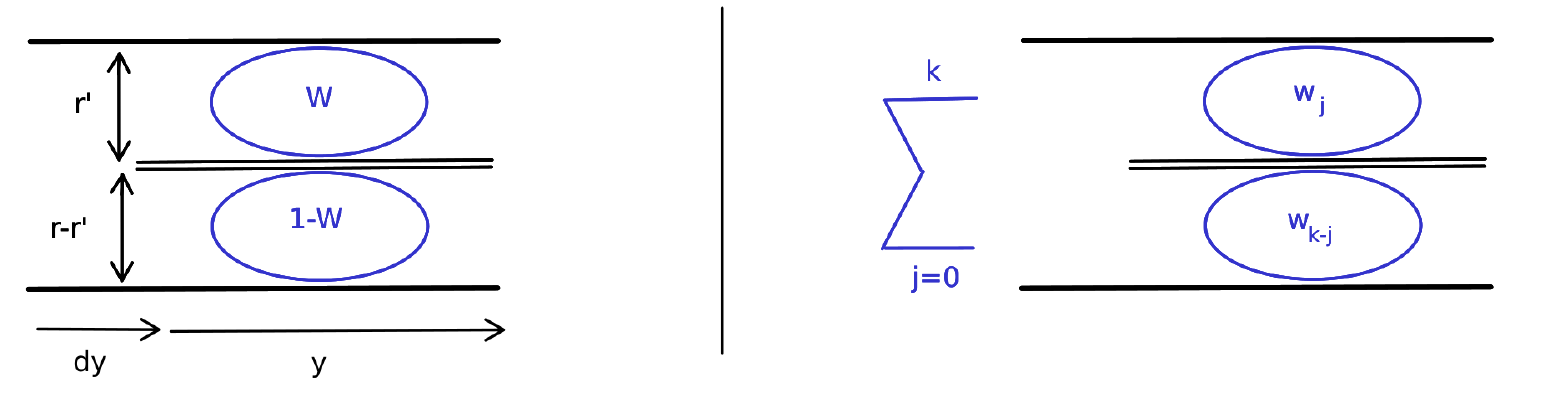}
  \end{center}
  \caption{\label{fig:proof-equations}\small
    Contributions of dipole splitting events to
    the evolution of $W$ (left; This is one of the two possible configurations,
    the other one being deduced from the one displayed through the
    exchange $W\leftrightarrow 1-W$), and of $w_k$ (right).
    }
\end{figure}

Using well-known techniques, we
may easily establish an evolution equation for $W$
with the total rapidity $y\geq y_0$.
This goes as follows: We start from the restframe of the onium, and increase
the total rapidity of the scattering process by $dy$ through a boost of the onium.
In the small rapidity interval $dy$, the initial dipole of transverse size $\underline{r}$
splits into a pair of dipoles of respective sizes
$\{\underline{r}',\underline{r}-\underline{r}'\}$
with probability $dy\, dp_{1\rightarrow 2}(\underline{r},\underline{r}')$,
or stays a single dipole with
probability $1-dy \int dp_{1\rightarrow 2}(\underline{r},\underline{r}')$.
One notices that if it splits, then the only contributions to $W(y+d y,r;y_0)$ come
from the following configurations: 
The further fluctuations of one offspring eventually scatter an odd number
of times, while the fluctuations of the other one
either scatter an even number of times or do not scatter at all.
The probability of such configurations reads
\be
W(y,r';y_0)\times \left[1-W(y,|\underline{r}-\underline{r}'|;y_0)\right]
+\left\{\underline{r}'\leftrightarrow \underline{r}-\underline{r}'\right\}.
\ee
The first term is illustrated in  Fig.~\ref{fig:proof-equations}.

Putting together the contributions of all possible cases
weighted by the corresponding probabilities, and letting $dy\rightarrow 0$, we get
the integro-differential equation
\begin{multline}
  \partial_y W(y,r;y_0)=\int_{\underline{r}'}dp_{1\rightarrow 2}(\underline{r},\underline{r}')
  \bigg[
  W(y,r';y_0)+W(y,|\underline{r}-\underline{r}'|;y_0)
  -2W(y,r';y_0)W(y,|\underline{r}-\underline{r}'|;y_0)\\
  -W(y,r;y_0)
  \bigg].
\label{eq:evolW}
\end{multline}
The three first terms under the integral in the right-hand side
come from the splitting events, the last term from non-splitting
events.
For $y=y_0$, the rapidity at which we want to take the initial condition,
the onium Fock state reduces to a single dipole of size~$r$:
$W$ then just equals the amplitude $T$ at $y=y_0$,
\be
W(y_0,r;y_0)=T(y_0,r).
\label{eq:initW}
\ee

Comparing Eq.~(\ref{eq:evolW}) with Eq.~(\ref{eq:BKT}),
we see that the function $2W$
obeys \BKT, but with as an initial
condition\footnote{%
As a side remark, $2T$
is a function that takes values between~0 and~2, while the stable fixed
point of Eq.~(\ref{eq:BKT}) is ${\mathfrak T}=1$, which is a bit unusual.}
$W(y_0,r;y_0)=2T(y_0,r)$,
namely the first term of the initial condition Eq.~(\ref{eq:sigmain-IC-KL})
for the KL equation. We note that this term is dominant
in the parametric region in which $T\ll 1$, and this allows us to argue that
the solutions of the KL equation on the one hand, and $2W$ on the other hand,
match asymptotically. Indeed, this stems from a general property of nonlinear
equations in the class
of the Fisher~\cite{Fisher.1937}
and Kolmogorov-Petrovsky-Piscounov~\cite{KPP.1937} (FKPP) equation,
to which the \BKT equation
belongs~\cite{Munier:2003vc}. The large-rapidity solutions of such equations
are essentially determined by the shape of the small-$r$ (i.e. large positive-$x$) tail
of the initial condition~\cite{VANSAARLOOS200329}. In this region, $2T-T^2\simeq 2T$:
the two functions taken as initial conditions coincide, and thus, the large-$y$
solutions are the same.
Therefore, for our purpose of deriving the exact asymptotics of
solutions to the KL equation,
we can safely trade $\sigma_\text{in}$ for twice the sum $W$ of all odd weights,
and consequently, $\sigma_\text{diff}$ for twice the sum of all even weights,
as in Eq.~(\ref{eq:sigmadiff}).

%%%%%%%%%%%%%%%%%%%%%%%%%%%%%%%%%%%%%%%%%%%%%%%%%%%%%%%%%%%%%%%%%%%%%%%%%%%%%%%%%%%%%%%%%
%%%%%%%%%%%%%%%%%%%%%%%%%%%%%%%%%%%%%%%%%%%%%%%%%%%%%%%%%%%%%%%%%%%%%%%%%%%%%%%%%%%%%%%%%

\section{Analytical asymptotics for the weights of the number of participating
  dipoles\label{sec:analytics}}

\subsection{Heuristic calculation}

In this section, we shall compute the $w_k$'s from their definition~(\ref{eq:def-w_k})
as the probability
that exactly $k$ dipoles in the Fock state of the onium at rapidity
$\tilde y_0$ interact with the nucleus boosted to the
rapidity $y_0$.
The main ingredient is the overlap integral~$I^{(1)}$ defined in Eq.~(\ref{eq:I-approx}).
$T$ in there is the forward elastic scattering amplitude of a dipole,
which solves the Balitsky-Kovchegov equation, with as an initial condition,
the scattering amplitude of a dipole
off a nucleus at zero rapidity: Its expression is derived from
Eq.~(\ref{eq:BK-solution}).

We will also need the form of the dipole number density $n$
in realizations of the QCD evolution,
endowed with its distribution,
in order to be able to take the expectation value
in Eq.~(\ref{eq:def-w_k}).
It is not known analytically: Therefore,
we shall use the phenomenological model for QCD evolution
to represent the dipole content of the onium.
This model was introduced in Ref.~\cite{Mueller:2014gpa} for general branching
random walks, and recently used in the context
of QCD in Ref.~\cite{Le:2020zpy}.
We start with a brief review of the phenomenological model,
before giving the expression of the overlap integral. We will
then be in a position to calculate the weights $w_k$.

\subsubsection{Formulation of the phenomenological model}

The main assumption of the phenomenological model
is that the onium Fock state evolves deterministically, except for
one single fluctuation consisting in one unusually large dipole produced
at some random rapidity $\tilde y_1$,
which subsequently to its emission, also evolves deterministically
until the scattering rapidity $\tilde y_0$.

The deterministic, or ``mean-field'', evolution of particles undergoing a branching random
walk process was studied e.g. in Ref.~\cite{Mueller:2014gpa}. It was shown that
the mean-field dipole number density of log inverse size $x'$
in an onium of initial log inverse size ${\mathfrak X}$,
after evolution over the rapidity interval $\Delta \tilde y$, reads
\be
\bar n(\Delta \tilde y,x^{\prime}-{\mathfrak X})=C_1(x'-{\mathfrak X}-\tilde X_{\Delta\tilde y})
\,e^{\gamma_0(x'-{\mathfrak X}-\tilde X_{\Delta\tilde y})}
\exp\left(-\frac{(x'-{\mathfrak X}-\tilde X_{\Delta\tilde y})^2}{2\chi''(\gamma_0)\Delta\tilde y}
\right)
\Theta(x'-{\mathfrak X}-\tilde X_{\Delta \tilde y}),
\label{eq:bar-n}
\ee
where $\tilde X_{\Delta\tilde y}=-\chi'(\gamma_0)\Delta\tilde y+\frac{3}{2\gamma_0}\ln\Delta\tilde y$.
(There may be an additive constant of order unity, but in Eq.~(\ref{eq:bar-n}),
it is either negligible compared to other terms, or, when exponentiated, absorbed into
the overall constant).

As for the distribution
of the rapidity and size of the fluctuation, we introduce the joint probability density
$p(\delta,\tilde y_1)$ that the log inverse size of the unusually-large
dipole be smaller by $\delta$
than the log inverse size of the typical largest dipole
${\mathfrak X}+\tilde X_{\tilde y_1}$, and that it occurs at rapidity $\tilde y_1$.
We assume that it coincides with the
distribution of the relative log inverse size of the largest dipole at $\tilde y_1$.
The probability that there is at least one dipole with size larger than some fixed size,
from which
one deduces $p(\delta,\tilde y_1)$ by simple derivation and change of variables,
solves the equation \BKT, with as an initial condition an appropriate Heaviside distribution.
Thus, for $\delta\gg 1$,
\be
p(\delta,\tilde y_1)=C\,\delta\, e^{-\gamma_0\delta}
\exp\left(-\frac{\delta^2}{2\chi''(\gamma_0)\tilde y_1}\right).
\label{eq:p}
\ee
(Again, $C$ is an undetermined numerical constant of order unity).

The number density of dipoles of log inverse size $x'$ at rapidity $\tilde y_0$,
starting with an onium of log inverse size $x$, reads
\be
n(x')=\bar n(\tilde y_0,x'-x)+\bar n(\tilde y_0-\tilde y_1,x'-x-\tilde X_{\tilde y_1}+\delta)
\label{eq:n-pheno}
\ee
with probability $p(\delta,\tilde y_1)\, d\delta\, d\tilde y_1$.
The first term in the right-hand side
is the deterministic evolution of the initial dipole of log inverse size
${\mathfrak X}\equiv x$. The second term
represents the particle density generated by the
evolution of the fluctuation of initial log inverse size
${\mathfrak X}\equiv x+\tilde X_{\tilde y_1}-\delta$.
When $x-X_y$ is
large, the former is necessarily small compared to the latter,
and therefore, can be neglected.

\subsubsection{Overlap}

The overlap $I$ of the particle number is given by Eq.~(\ref{eq:I-approx}),
up to the replacement of $n$ by the second term in Eq.~(\ref{eq:n-pheno}),
and of $T$ by  Eq.~(\ref{eq:BK-solution}) with ${\mathfrak T}=T$. (We will
call the overall numerical constant appearing in that formula $C_2$).
Thus, in the phenomenological model, it reads
\be
I^{(1)}_{\delta,\tilde y_1}(y,x;y_0)=\int dx'\, \bar n(\tilde y_0-\tilde y_1,x'-x-\tilde X_{\tilde y_1}
+\delta)\,T(y_0,x').
\ee

Now we consider a fluctuation of size $\delta$ occuring at
the rapidity $\tilde y_1<\tilde y_0$ such that
$\tilde y_0-\tilde y_1\gg 1$.
We choose $\delta$ so that $x-X_y-\delta$ be positive, large compared to unity
but small compared to $\sqrt{y_0}$, which is always possible if $y_0$ is large enough.
This restriction is motivated by the fact that configurations in this class, which maximize
the overlap integral $I^{(1)}$ multiplied by the probability of
a fluctuation $p(\delta,\tilde y_1)$,
turn out to be the dominant ones when computing the averages over the onium Fock states;
see Ref.~\cite{Le:2020zpy}.
The calculation of $I^{(1)}$, rather straightforward in the regime of interest,
was performed in the previous reference, and led to the following expression:
\be
I^{(1)}_{\delta,\tilde y_1}(y,x;y_0)
=C_1C_2\sqrt{\frac{\pi}{2}}\left[\chi''(\gamma_0)\right]^{3/2}e^{-\gamma_0(x-X_y)}
\left(\frac{y}{y_1\tilde y_1}
\right)^{3/2}
e^{\gamma_0\delta}.
\label{eq:I0}
\ee
This formula is manifestly boost invariant, since it does not exhibit a dependence on $y_0$.
Therefore, in what follows,
we shall cancel $y_0$ from the list of variables upon which $I^{(1)}$ depends.

\subsubsection{Evaluation of the weights $w_k$}

We are now ready to calculate the weights $w_k$. In the phenomenological model, 
this calculation is formulated as
\be
w_k(y,x;y_0)\equiv\left\langle
\frac{1}{k!}\left[I^{(1)}_{\delta,\tilde y_1}(y,x)\right]^k
e^{-I^{(1)}_{\delta,\tilde y_1}(y,x)}
\right\rangle_{\delta,\tilde y_1\leq\tilde y_0}
\ee
where the average $\langle\cdot\rangle$ over the Fock states takes the form
of an integration over 
the rapidity $\tilde y_1$ at which the fluctuation may occur, and over its size $\delta$,
weighted by their probability distribution $p(\delta,\tilde y_1)$.
In other terms,
\be
w_k(y,x;y_0)=\int_0^{\tilde y_0}d\tilde y_1\int_0^{+\infty}
d\delta\,p(\delta,\tilde y_1)
\frac{1}{k!}\left[I^{(1)}_{\delta,\tilde y_1}(y,x)\right]^k
e^{-I^{(1)}_{\delta,\tilde y_1}(y,x)}.
\ee

We start by computing the integral over $\delta$ only: This is just
the density of the $\tilde y_1$-variable.
After replacing $p$ by its expression given in Eq.~(\ref{eq:p}),
using $I^{(1)}_{\delta,\tilde y_1}$ (denoted $I$)
as an integration variable instead of $\delta$, we write
\begin{multline}
\frac{\partial w_k}{\partial \tilde y_1}
(y,x;y-\tilde y_1)%\right|_{y_0=y_1}
=\frac{C}{\gamma_0^2}
\frac{1}{k!}I^{(1)}_{0,\tilde y_1}(y,x)
\int_{I^{(1)}_{0,\tilde y_1}(y,x)}^\infty
  dI\,\ln\frac{I}{I^{(1)}_{0,\tilde y_1}(y,x)}\,I^{k-2}e^{-I}\times\\
  \times\exp
  \left(-\frac{\ln^2 \left[{I}/{I^{(1)}_{0,\tilde y_1}(y,x)}\right]}
       {2\gamma_0^2\chi''(\gamma_0)\tilde y_1}
  \right).
\end{multline}
Integrals of this form also appeared in Ref.~\cite{Le:2020zpy}: It was argued that the last
exponential can be replaced by~1, and the remaining integral could be performed in the
relevant limit (see appendix~A in the latter paper).
Here we improve the treatment of this integral:
We write the last exponential in the form of a series,
\be
  %-\left.
  \frac{\partial w_k}{\partial \tilde y_1}(y,x;y-\tilde y_1)%\right|_{y_0=y_1}
  =\frac{C}{\gamma_0^2}
\frac{1}{k!}I^{(1)}_{0,\tilde y_1}(y,x)
\sum_{n=0}^\infty\frac{1}{n!}
\frac{(-1)^n}{[2\gamma_0^2\chi''(\gamma_0)\tilde y_1]^n}
\int_{I^{(1)}_{0,\tilde y_1}(y,x)}^\infty
\frac{dI}{I}\,\ln^{2n+1}\left(\frac{I}{I^{(1)}_{0,\tilde y_1}(y,x)}\right)
\,I^{k-1}\,e^{-I}.
 \label{eq:w_k-start}
\ee
Note that for any $k\geq 2$, the integral is regular when 
$I^{(1)}_{0,\tilde y_1}\rightarrow 0$.
But it is logarithmically divergent in the case $k=1$, which will require
a separate treatment.

We are now going to compute the leading terms in the joint large-$\tilde y_1$,
large-$|\ln I^{(1)}_{0,\tilde y_1}|$ (i.e. large $x-X_y$)
limit. We will keep only the leading power of $I^{(1)}_{0,\tilde y_1}$,
and resum the terms in the series
which possess the maximum number of 
factors $|\ln I^{(1)}_{0,\tilde y_1}|\sim \gamma_0(x-X_y)$
for each power of
$1/\sqrt{\tilde y_1}$. We shall first address the values of $k$ larger
than or equal to~$2$,
which can all be treated in the same way, and then address
the case $k=1$.

\paragraph{Case $k\geq 2$.}
It is useful to represent the logarithms that appear in Eq.~(\ref{eq:w_k-start})
by derivatives of power functions. Then, the integral therein reads
\begin{multline}
\int_{I^{(1)}_{0,\tilde y_1}(y,x)}^\infty
\frac{dI}{I}\,\ln^{2n+1}\left(\frac{I}{I^{(1)}_{0,\tilde y_1}(y,x)}\right)
\,I^{k-1}\,e^{-I}\\
=\left.\frac{\partial^{2n+1}}{\partial\alpha^{2n+1}}\right|_{\alpha=0}
\left\{\left[I^{(1)}_{0,\tilde y_1}(y,x)\right]^{-\alpha}
\times\Gamma[\alpha+k-1,I^{(1)}_{0,\tilde y_1}(y,x)]\right\},
\label{eq:log-derivative-rep}
\end{multline}
where $\Gamma$ is the incomplete Euler-Gamma function
\be
\Gamma(x,I)\equiv\int_{I}^\infty
d\bar I\,{\bar I}^{x-1}\,e^{-\bar I}.
\ee
The leading log terms are obtained when all
the derivatives with respect to $\alpha$ act on the factor
$[I^{(1)}_{0,\tilde y_1}(y,x)]^{-\alpha}$, and the incomplete
Gamma function is replaced by the complete one, which evaluates as the factorial $[(k-2)!]$.
After trivial simplifications, we find
\be
\frac{\partial w_k}{\partial \tilde y_1}(y,x;y-\tilde y_1)
=\frac{C}{\gamma_0^2}
\frac{1}{k(k-1)}I^{(1)}_{0,\tilde y_1}\,\ln\frac{1}{I^{(1)}_{0,\tilde y_1}}\,
\sum_{n=0}^\infty\frac{(-1)^n}{n!}
\frac{\ln^{2n} I^{(1)}_{0,\tilde y_1}}{[2\gamma_0^2\chi''(\gamma_0)\tilde y_1]^n},
\ee
where we have understood the variables upon which the function $I^{(1)}_{0,\tilde y_1}$
depends.

The leading-log series can easily be resummed in the form of an exponential.
Replacing $I^{(1)}_{0,\tilde y_1}$ by its expression~(\ref{eq:I0})
and $(-\ln I^{(1)}_{0,\tilde y_1})$ by $\gamma_0(x-X_y)$,
which amounts to neglecting constants and logarithms of $y$, $y_1$, $\tilde y_1$
compared to $x-X_y$, we get
\be
  \frac{\partial w_k}{\partial \tilde y_1}(y,x;y-\tilde y_1)
=
\frac{c}{\gamma_0}\frac{1}{\sqrt{2\pi\chi''(\gamma_0)}}\frac{1}{k(k-1)}
(x-X_y)\,e^{-\gamma_0(x-X_y)}
\left(\frac{y}{y_1\tilde y_1}
\right)^{3/2}
\exp\left(
-\frac{(x-X_y)^2}{2\chi''(\gamma_0)\tilde y_1}
\right),
\label{eq:w_k.ge.2}
\ee
where we have defined the constant $c$ as a product of the previously-introduced
undetermined numerical constants,
\be
c\equiv {CC_1C_2}\,
\pi\left[\chi''(\gamma_0)\right]^{2}.
\ee

The integration over the rapidity $\tilde y_1$ at which the
fluctuation occurs, 
\be
w_k(y,x;y_0)=\int_0^{\tilde y_0}d\tilde y_1\,\partial_{\tilde y_1}w_k(y,x;y-\tilde y_1),
\ee
is well-defined, due to the last exponential in Eq.~(\ref{eq:w_k.ge.2})
which acts as a diffusive cutoff on small values of
$\tilde y_1$.
We may write the integral with the help of an error function
and of elementary functions as
\begin{multline}
\int_0^{\tilde y_0}d\tilde y_1
\left(\frac{y}{y_1\tilde y_1}
\right)^{3/2}
\exp\left(
-\frac{(x-X_y)^2}{2\chi''(\gamma_0)\tilde y_1}
\right)\\
=
  \frac{\sqrt{2\pi\chi''(\gamma_0)}}{x-X_y}
  \left(
  1-\frac{(x-X_y)^2}{\chi''(\gamma_0)y}
  \right)
  \text{erfc}\left(
  \frac{x-X_y}{\sqrt{2\chi''(\gamma_0)}}\sqrt{\frac{y_0}{y\tilde y_0}}
  \right)\exp\left(-\frac{(x-X_y)^2}{2\chi''(\gamma_0)y}\right)\\
  +2\sqrt{\frac{\tilde y_0}{y y_0}}
  \exp\left(-\frac{(x-X_y)^2}{2\chi''(\gamma_0)\tilde y_0}\right).
\label{eq:exact}
\end{multline}
When $y\rightarrow\infty$, it boils down to two simple terms.
Therefore, in this limit, $w_k$ eventually reads
\be
w_{k\geq 2}(y,x;y_0)=\frac{c}{\gamma_0}\frac{1}{k(k-1)}
\left(
1+\sqrt{\frac{2}{\pi\chi''(\gamma_0)}}\frac{x-X_y}{\sqrt{y_0}}
\right)e^{-\gamma_0(x-X_y)}.
\label{eq:w_k-final}
\ee
Interestingly enough, the ratio $w_{k\geq 2}/w_2$ has a very simple expression:
\be
\frac{w_{k\geq 2}}{w_2}
=\frac{2}{k(k-1)}.
\ee
This shows that the distribution of the number of participating dipoles
decreases only slowly at large~$k$.
The events which involve many of them are not rare at all.
As a matter of fact, the mean participant number is formally infinite.

%%%%%

\paragraph{Case $k=1$.}
In this case, it is convenient to change variable
in the integral in~Eq.~(\ref{eq:w_k-start}). We define
$l\equiv\ln [I/I^{(1)}_{0,\tilde y_1}(y,x)]$: the integral over $I$ then
becomes
\be
\int_0^{+\infty} dl\,l^{2n+1}\exp\left({-I^{(1)}_{0,\tilde y_1}(y,x)\,e^{l}}\right).
\ee
The exponential is tantamount to a cutoff effectively limiting
the integration region to $[0,-\ln I^{(1)}_{0,\tilde y_1}(y,x)]$.
For small $I^{(1)}_{0,\tilde y_1}$,
we can replace it by a Heaviside-theta function, and hence perform
trivially the integral. We get
\be
\frac{1}{2n+2}\ln^{2n+2}I^{(1)}_{0,\tilde y_1}(y,x).
\ee
After resummation of the series of the
leading logarithms of $I^{(1)}_{0,\tilde y_0}$ and simplifications along the
same lines as those followed in the case $k\geq 2$, one finds
\be
%-\left.
\frac{\partial w_1}{\partial \tilde y_1}(y,x;y-\tilde y_1)=
%\right|_{y_0=y_1}
{c}{\sqrt{\frac{\chi''(\gamma_0)}{2\pi}}}
\,e^{-\gamma_0(x-X_y)}
\frac{y^{3/2}}{y_1^{3/2}\sqrt{\tilde y_1}}
\left[1-
\exp\left(
-\frac{(x-X_y)^2}{2\chi''(\gamma_0)\tilde y_1}
\right)
\right].
\ee

As in the case $k\geq 2$, we may integrate
over the rapidity $\tilde y_1$ between $0$ and $\tilde y_0$.
Now the singularity at $\tilde y_1=0$ is not cut off, but it is integrable.
Since we assume that $\tilde y_0$ is on the order of $y$, and since
we take $x$ in the scaling
region such that $(x-X_y)^2\ll y$, the upper limit of the relevant integration region
is on the order of $(x-X_y)^2$.
We can approximate $(y/y_1)^{3/2}$ by $1$, and set the upper bound to $+\infty$.
The integral to perform takes the form
\be
\int_0^\infty \frac{d\tilde y_1}{\sqrt{\tilde y_1}}\left[1-
\exp\left(
-\frac{(x-X_y)^2}{2\chi''(\gamma_0)\tilde y_1}
\right)
\right]\simeq \sqrt{\frac{2\pi}{\chi''(\gamma_0)}}\times (x-X_y).
\ee
Hence, the weight of the contributions of a single participating dipole reads
\be
w_1(y,x;y_0)=c(x-X_y)\times e^{-\gamma_0(x-X_y)}.
\label{eq:w_1-final}
\ee
At this level of approximation, $w_1$ is manifestly boost invariant.
Actually, what is rigorously boost invariant is the total cross section $\sigma_\text{tot}$,
i.e. the series $2\sum_{k\geq 1} w_k$. But the term $2w_1$ dominates this sum
parametrically,
since it has an extra $x-X_y$ factor with respect to all other terms.

%%%%%%%%%%%%%%%%%%%%%%%%%%%%%%%%%%%%%%%%%%%%%%%%%%%%%%%%%

\subsection{Generating function\label{sec:generating-function}}

In this section, we will establish that a generating function of the weights $w_k$
obeys a set of BK equations. We will conjecture the large-rapidity solution, and check it
numerically.

\subsubsection{Rapidity evolution of the weights and of their generating
function}

The set of weights $\{w_k,k\geq 0\}$
obeys a hierarchy of evolution equations,
\be
\partial_y w_k(y,r;y_0)=\int_{\underline{r}'}dp_{1\rightarrow 2}(\underline{r},\underline{r}')
\left(\sum_{j=0}^kw_j(y,{r}';y_0)w_{k-j}(y,|\underline{r}-\underline{r}'|;y_0)
-w_k(y,r;y_0)
  \right),
\ee
with the initial condition
$w_k(y_0,r;y_0)=\delta_{k,0}S(y_0,r)+\delta_{k,1}T(y_0,r)$. The proof is straightforward,
using well-known techniques;
See Fig.~\ref{fig:proof-equations} for a graphical illustration of the contribution
of the non-trivial first term in the right-hand side.

The generating function
\be
\tilde w_\lambda(y,r;y_0)=\sum_{k=0}^\infty \lambda^k\,w_k(y,r;y_0)
\label{eq:generating-power-series}
\ee
obeys a unique equation, which turns out to be
the BK${}_{\mathfrak S}$ equation~(\ref{eq:BKS}), with ${\mathfrak S}\equiv \tilde w$.
The initial condition at $y=y_0$ reads
\be
\tilde w_\lambda(y_0,r;y_0)=1-(1-\lambda)T(y_0,r).
\label{eq:ICw}
\ee

A few comments are in order.
First, the unitarity of the probability that any scattering may occur
  reads, in terms of the generating function, $\tilde w_{\lambda=1}(y,r;y_0)=1$.
Second, the $S$-matrix coincides with the generating function evaluated at $\lambda=0$:
  $\tilde w_{\lambda=0}(y,r;y_0)=S(y,r)$.
Third, there is a direct relation
between the generating function evaluated at two different values of~$\lambda$
and the difference of the diffractive and
inelastic cross sections:
\be
2\left(\tilde w_{\lambda=-1}-\tilde w_{\lambda=0}\right)=\sigma_\text{diff}-\sigma_\text{in}.
\ee

%%%%%%%%%%%%%%%%%%%%%%%%%%%%%%%%%%%%%%%%%%%%%%%%%%%%%%%%%%%%

\subsubsection{Infinite-rapidity limit: traveling wave solution}

In the infinite-$y$ limit, the solution
to the BK equation converges to a traveling wave also starting with
the initial condition~(\ref{eq:ICw}), namely that
$\tilde w_\lambda(y,x;y_0)$ tends to a function of $x-X_y+f_{y_0}(\lambda)$
only, where $X_y$ was given in Eq.~(\ref{eq:Xy}),
and $f$ is a ``delay function'' that vanishes for $\lambda=0$.

When furthermore $x-X_y+f_{y_0}(\lambda)$ is taken finite but large,
the analytic form for the shape of the traveling wave front reads
\be
1-\tilde w_\lambda(y,x;y_0)=c\,
\left[x-X_y+f_{y_0}(\lambda)\right]
e^{-\gamma_0[x-X_y+f_{y_0}(\lambda)]},
\label{eq:traveling-wave}
\ee
where $c$ is an undetermined constant of order unity.

As for the delay function $f_{y_0}(\lambda)$,
we may conjecture the formula
\be
f_{y_0}(\lambda)=\frac{1}{\gamma_0}\ln\frac{1}{1-\lambda}
\times\left(
1-\frac{1}{\gamma_0}\sqrt{\frac{2}{\pi\chi''(\gamma_0)}}
  \frac{1}{\sqrt{y_0}}
  \right).
\label{eq:conjecture-delay}
\ee
This conjecture is motivated by the following observation.
There exists a function $f_{y_0}(\lambda)$ such
that the expression for
$1-\tilde w_\lambda(y_0,x;y_0)$ in Eq.~(\ref{eq:ICw}), namely
\be
c(1-\lambda)(x-X_{y_0})e^{-\gamma_0(x-X_{y_0})}
\ee
can be matched to the regular (delayed) traveling wave~(\ref{eq:traveling-wave})
at $y=y_0$, in the region $\ln\frac{1}{1-\lambda}\ll x-X_{y_0}\ll\sqrt{y_0}$.
Indeed, equalizing the two front shapes implies
\be
e^{\gamma_0 f_{y_0}(\lambda)}=\frac{1}{1-\lambda}\times\left(1+\frac{f_{y_0}(\lambda)}{\Delta}\right)
\ee
which can be solved iteratively as
\be
f_{y_0}(\lambda)=\frac{1}{\gamma_0}\ln\frac{1}{1-\lambda}
+\frac{1}{\gamma_0}\ln\left(1
+\frac{\frac{1}{\gamma_0}\ln\frac{1}{1-\lambda}+\frac{1}{\gamma_0}
\ln\left(1+\frac{f_{y_0}(\lambda)}{\Delta}\right)}{\Delta}\right).
\ee
For $\Delta$ satisfying the above ordering condition,
the second logarithm may be expanded to first order, leading to a
closed expression for the delay:
\be
f_{y_0}(\lambda)\simeq\frac{1}{\gamma_0}\ln\frac{1}{1-\lambda}
\left(1+\frac{1}{\gamma_0\Delta}\right).
\ee
Keeping $\lambda$ fixed, choosing $\Delta$
of order say $y_0^{1/2-\epsilon}$ for some fixed $\epsilon\in]0,\frac12[$
and letting $y_0$ become large,
this expression clearly tends to the conjectured $f_{\infty}(\lambda)$,
Eq.~(\ref{eq:conjecture-delay}).
Since the expression for $1-\tilde w_\lambda(y_0,x;y_0)$
is that of a regular
front which would have evolved from a step initial condition $\Theta(-x)$
in a large region from its tip,
and since the large-rapidity position of a traveling wave is 
determined precisely by its shape in the tip region,
we conclude that the solution for
$1-\tilde w_\lambda(y,x;y_0)$ indeed tends to a traveling wave
at large~$y$, with position
$X_y$ pulled back by the distance $f_{y_0}(\lambda)$.
We expect this solution to be valid when $|\ln(1-\lambda)|\ll \sqrt{y_0}$,
which is not a too restrictive condition,\footnote{%
  This condition limits the values of $k$ we may reach through our calculation
  to numbers much smaller than ${\cal O}(e^{\text{const}\times\sqrt{y_0}})$;
  but this is parametrically
  a very large number when ${y_0}\gg 1$.
}
since we are eventually
interested in the expansion of
$\tilde w_\lambda$ around $\lambda=0$.
We are not able to determine the finite-$y_0$ correction from
these heuristics, but since
$\Delta$ is at most $\sqrt{y_0}$, a subleading term of relative
order $1/\sqrt{y_0}$ is plausible.

Note that $f_{\infty}(\lambda)$
is the known leading contribution to the delay when the initial condition for
the \BKT equation (or for any equation in the same class)
is a step function of height
$1-\lambda$~\cite{Brunet_2009,Mueller:2019ror}.
But in that case, $\frac{1}{\gamma_0}\ln\frac{1}{1-\lambda}$
represents the largest term in the expression of the
delay in the limit $\lambda\rightarrow 1$,
and there is a subleading non-analytic term
$-\frac{1}{\gamma_0}\ln[-\ln(1-\lambda)]$.\\

The numerical coefficient of the subleading term is
chosen in order to recover the expressions for $w_k$
found above in the phenomenological model. Indeed,
the shape of the traveling wave~(\ref{eq:traveling-wave}) with
$f_{y_0}(\lambda)$ being replaced by Eq.~(\ref{eq:conjecture-delay})
can be expanded in the power series~(\ref{eq:generating-power-series}) of $\lambda$,
leading to expressions for the
$w_k$'s. Let us outline the main steps of the calculation.

Starting from Eq.~(\ref{eq:traveling-wave}), replacing $f$ in there by
Eq.~(\ref{eq:conjecture-delay}), we get
\be
1-\tilde w_\lambda(y,x;y_0)=c(1-\lambda)^{1-\frac{1}{\gamma_0}
  \sqrt{\frac{2}{\pi\chi''(\gamma_0)}}\frac{1}{\sqrt{y_0}}}
\left[
  x-X_y+\frac{1}{\gamma_0}\ln\frac{1}{1-\lambda}\left(
  1-\frac{1}{\gamma_0}
  \sqrt{\frac{2}{\pi\chi''(\gamma_0)}}\frac{1}{\sqrt{y_0}}
  \right)
  \right]e^{-\gamma_0(x-X_y)}.
\ee
Next, we expand for large $y_0$,
dropping all higher powers
of $1/\sqrt{y_0}$, and the terms of order $1/\sqrt{y_0}$
which are not enhanced by a power of $x-X_y$. The generating function
then reads
\be
1-\tilde w_\lambda(y,x;y_0)=\left[c(1-\lambda)(x-X_y)
+\frac{c}{\gamma_0}(1-\lambda)\ln\frac{1}{1-\lambda}
\left(
1+\sqrt{\frac{2}{\pi\chi''(\gamma_0)}}\frac{x-X_y}{\sqrt{y_0}}
\right)\right]e^{-\gamma_0(x-X_y)}.
\ee
Finally, we expand in power series of $\lambda$,
using the identity
\be
(1-\lambda)\ln\frac{1}{1-\lambda}=\lambda-\sum_{k\geq 2} \frac{\lambda^k}{k(k-1)}.
\ee
It is then straightforward to check that we get back Eq.~(\ref{eq:w_1-final})
for the coefficient $w_1$ of $(-\lambda)$ in this series (in the same approximations),
and Eq.~(\ref{eq:w_k-final}) for the coefficient $w_{k}$ of $(-\lambda^{k})$
in the case $k\geq 2$.\\

Let us comment that the proposed conjecture does not only apply to the
present context, but applies also much more generally
to a large class of branching random walk models.
This allows for accurate checks: Indeed, we can pick
a model easy to implement numerically and to run, and solve it for the delay.
Such a calculation is reported in the appendix and shows perfect consistency
with our conjecture.
The good matching of our numerical calculation with the
conjecture brings, in turn, strong support for the
expressions of the weights $w_k$ we have found from the phenomenological
model, since they are fully determined by
the delay of the traveling wave in the present approach.

%%%%%%%%%%%%%%%%%%%%%

\section{Diffractive cross section and gap distribution\label{sec:diff}}

We are now in a position to come back to the physical observables, and
with the help of the results we have obtained in the previous section,
provide asymptotic expressions. We shall then discuss the problems posed by a standard
perturbative formulation.

\subsection{Analytical asymptotics}

As argued above,
we get the correct parametric expression for $\sigma_\text{tot}$ by identifying
it to $2w_1$.

As for the diffractive cross section,
we get it by summing $2w_k$ over the even values of $k$, starting with
$k=2$. Using Eq.~(\ref{eq:w_k-final}) for $w_{k\geq 2}$ and Eq.~(\ref{eq:w_1-final})
for $w_1$, we arrive at a very simple expression:
\be
\frac{\sigma_{\text{diff}}(y,x;y_0)}{\sigma_{\text{tot}}(y,x)}=\frac{\ln 2}{\gamma_0}
\left(\frac{1}{x-X_y}
  +\sqrt{\frac{2}{\pi\chi''(\gamma_0)}}\frac{1}{\sqrt{y_0}}\right).
\label{eq:diff-final}
\ee
To get this result, we just needed to use the trivial identity
\be
\sum_{\text{even}\ k\geq 2}\frac{1}{k(k-1)}=\ln 2.
\label{eq:ln2}
\ee
This formula for $\sigma_\text{diff}/\sigma_\text{tot}$
is expected to be valid asymptotically for large $y$ and large $y_0$,
and for $x$, chosen in the scaling region, i.e. such that
$(x-X_y)^2 \ll {y}$.

Let us interpret the two terms in Eq.~(\ref{eq:diff-final}).
The fluctuation creating a large dipole, that scatters elastically off the nucleus,
happens most likely either in the beginning of the evolution
(leading to a dissociative but small mass event)
or close to the scattering rapidity $\tilde y_0$ (leading to a gap of size close to $y_0$).
The first configuration is dominant when $y_0$ is chosen large compared to
$(x-X_y)^2$,
leading to the first term in Eq.~(\ref{eq:diff-final}).
The second configuration is dominant when $y_0$ is chosen
small compared to $(x-X_y)^2$: In this
case, the second, $y_0$-dependent, term dominates the diffractive cross section.

In the same way, starting from Eqs.~(\ref{eq:def-pi}),~(\ref{eq:sigmadiff}) and
using the analytical expression~(\ref{eq:w_k.ge.2}),
we find that the rapidity gap distribution in the scaling region reads
\be
\pi(y,x;y_\text{gap})=
\frac{\ln 2}{\gamma_0\sqrt{2\pi\chi''(\gamma_0)}}
\left(\frac{y}{y_\text{gap}(y-y_\text{gap})}\right)^{3/2}
\exp\left({-\frac{(x-X_y)^2}{2\chi''(\gamma_0)(y-y_\text{gap})}}\right).
\label{eq:pi}
\ee
Equation~(\ref{eq:diff-final}) is an integral of Eq.~(\ref{eq:pi}) over $y_\text{gap}$,
in the limit $x,y\rightarrow +\infty$, keeping $x-X_y$ fixed.
The determination of the overall constant is the main new result in this work,
while the functional dependence was first found in Ref.~\cite{Mueller:2018zwx}.

This expression is tantamount to that of the
distribution of the splitting rapidity of the most recent
common ancestor of the set of dipoles which scatter
derived in Ref.~\cite{Le:2020zpy}, up to an extra factor $\ln 2$.
This distribution,
denoted by $G(y,x;y_1)/T(y,x)$ in the latter paper, can be expressed with the
help of our weights $w_k$ as
\be
\frac{G(y,x;y_1)}{T(y,x)}=\frac{\sum_{k=2}^\infty \partial_{\tilde y_1}w_k(y,x;y-\tilde y_1)}
{\sum_{k=1}^\infty w_k(y,x;y_1)}.
\ee
Inserting the expression~(\ref{eq:w_k.ge.2})
for the $\partial_{\tilde y_1} w_k$'s into the numerator and replacing the
denominator by its leading term $w_1$ given in Eq.~(\ref{eq:w_1-final}),
we recover the analytical form found in Ref.~\cite{Le:2020zpy}.
The factor $\ln 2$ present in Eq.~(\ref{eq:pi}) does not appear here, basically
because the summation~(\ref{eq:ln2}) to arrive
at the gap distribution is
replaced by $\sum_{k\geq 2}1/[k(k-1)]$, which is unity.

%%%%%%%%%%%%%%%%%%%%%%%%%%%%%%%%%%%%%%%%%%%%%%%%%%%%%%%%%%%%%%%%%%%%%%%%%%%%%%%%%%%%%%%%%

\subsection{On the relation to the standard perturbative approach\label{sec:AGK}}

The formulation of the observables we are considering has involved the
probabilistic weights $w_k=\langle G_k\rangle$ of the contribution of
$k$ participant dipoles exactly. These quantities are probabilities in the dipole model,
but they have no simple diagrammatic interpretation: Rather, they resum an infinity
of diagrams, involving an arbitrary number of participating dipoles.

The standard approach consists instead in computing forward elastic
scattering amplitudes in perturbation theory, order by order in the number
of rescatterings. The contribution of each graph
to the considered observable is then obtained by applying the Abramovsky-Gribov-Kancheli
(AGK) cutting rules~\cite{Abramovsky:1973fm}.
The latter were initially established in the context of Regge theory,
and argued to also hold in QCD~\cite{Bartels:1996hw}.
The KL equation was proved to be consistent with these rules~\cite{Kovchegov:1999ji}.

In the formalism we have used in this paper, this approach corresponds
to expanding the observables as series of $\langle F_N(I)\rangle$, where
the $F_N$'s were defined in Eq.~(\ref{eq:FN_Gk}).
Indeed, one may write
\be
\sigma_\text{tot}(y,x)=2\sum_{N=1}^\infty (-1)^{N+1}\langle F_N[I(y_0)]\rangle_{\tilde y_0,x},\quad
\sigma_\text{in}(y,x)=\sum_{N=1}^\infty (-2)^{N+1}\langle F_N[I(y_0)]\rangle_{\tilde y_0,x},
\label{eq:F-expansion}
\ee
the diffractive cross section following from
the difference of the total and the inelastic cross
sections. (See also Ref.~\cite{Mueller:1996te} for a discussion of these
formulas starting from the AGK rules).
Note that the terms of order $N$ of these series are related to the $N$-th derivative
of the generating function $\tilde w_\lambda$, but evaluated at $\lambda=1$ instead
of $\lambda=0$ as in the case of the $w_k$'s.

But as shown in Ref.~\cite{Mueller:1994gb,Salam:1995uy},
the expansions~(\ref{eq:F-expansion})
turn out to be impractical to compute observables such as the
total or diffractive cross sections in the scaling region, where
unitarity corrections are important: Indeed, they were shown to be severely
divergent asymptotic series, and a Borel resummation is required to arrive
at a meaningful result.

%%%%%%%%%%%%%%%%%%%%%%%%%%%%%%%%%%%%%%%%%%%%%%%%%%%%%%%%%%%%%%%%%%%%%%%%%%%%%%%%%%%%%%%%
%%%%%%%%%%%%%%%%%%%%%%%%%%%%%%%%%%%%%%%%%%%%%%%%%%%%%%%%%%%%%%%%%%%%%%%%%%%%%%%%%%%%%%%%

\section{Conclusion and outlook\label{sec:conclusion}}

To summarize, we have found a purely probabilistic formulation
of diffractive onium-nucleus
scattering, that we expect to hold rigorously for large-enough values
of the rapidities. The very existence of such a formulation is already
surprising enough, since diffraction is a typical quantum mechanical
phenomenon, with no classical counterpart.

This formulation has enabled us to derive a parameter-free expression for
the ratio of the diffractive cross section with a fixed minimum
rapidity gap to the total
cross section, in the geometric scaling region and for large rapidities,
as well as for the gap distribution. In variables relevant to the scattering
of an onium of size $r$ off a large nucleus at relative rapidity $Y$,
this ratio reads~(see Eq.~(\ref{eq:diff-final}))
\be
\frac{\sigma_{\text{diff}}(Y,r;Y_0)}{\sigma_{\text{tot}}(Y,r)}=\frac{\ln 2}{\gamma_0}
\left(\frac{1}{2\ln [{1}/{rQ_s(Y)}]}
  +\sqrt{\frac{2}{\pi\chi''(\gamma_0)}}\frac{1}{\sqrt{\bar\alpha Y_0}}\right),
\label{eq:diff-final-conclusion}
\ee
where the $Y$-dependent saturation scale in units of the McLerran-Venugopalan
momentum evolves as
$\ln [Q_s^2(Y)/Q_A^2]=\chi'(\gamma_0) \bar\alpha Y -3/(2\gamma_0)\,\ln(\bar\alpha Y)$.
The gap distribution reads~(see Eq.~(\ref{eq:pi}))
\be
\pi(Y,r;Y_\text{gap})=
\frac{\ln 2}{\gamma_0\sqrt{2\pi\chi''(\gamma_0)}}
\frac{1}{\sqrt{\bar\alpha}}
\left(\frac{Y}{Y_\text{gap}(Y-Y_\text{gap})}\right)^{3/2}
\exp\left(-\frac{\ln^2 [r^2Q_s^2(Y)]}{2\chi''(\gamma_0)\bar\alpha (Y-Y_\text{gap})}\right).
\label{eq:pi-conclusion}
\ee

Along the way, we have found that diffraction is mostly due to the
exchange of a large
number of color singlets (``Pomerons'' in the language of Regge theory)
between the onium and the nucleon.
Indeed, the distribution $w_k$ of the number $k$ of
participant dipoles turns out to go like $1/[k(k-1)]$.

We believe that the present work paves the way for a number of
future developments. On the formal side,
a more rigorous derivation of the weights $w_k$ for general branching random
walks would be of great interest,
since the latter processes and their observables are
of potential relevance to different fields of science.
The generating function method exposed in Sec.~\ref{sec:generating-function}
looks a promising starting point.
Also, finding a systematic way to compute the subleading corrections would
be extremely useful. Knowing the next-to-leading order corrections,
presumably of relative
order $\ln \bar\alpha Y/\sqrt{\bar\alpha Y}$ or $1/\sqrt{\bar\alpha Y}$, would already
enable us to extend sizably the kinematical
range in which the asymptotic formulas are close to an exact calculation.

Finally, diffraction in onium-nucleus scattering can easily be related to the same
process in electron-ion collisions.
But approaching closely our analytical results would require
rapidities that are not reachable at colliders.
Nevertheless, since Eqs.~(\ref{eq:diff-final-conclusion}) and~(\ref{eq:pi-conclusion})
should represent the exact asymptotics of the KL equation, which follows
from QCD, they may be regarded as
a solid theoretical starting point for the construction of a realistic model
for diffractive dissociation, built in such a way that it matches, in the
appropriate limits, the asymptotics we have found.

%%%%%%%%%%%%%%%%%%%%%%%%%%%%%%%%%%%%%%%%%%%%%%%%%%%%%%%%%%%%%%%%%%%%%%%%%%%%%%%%%%%%%%%%

\section*{Acknowledgements}
The work of ADL and SM is supported in part by the Agence Nationale
de la Recherche under the project ANR-16-CE31-0019.
The  work  of  AHM  is  supported in part by
the U.S. Department of Energy Grant
DE-FG02-92ER40699.

%%%%%%%%%%%%%%%%%%%%%%%%%%%%%%%%%%%%%%%%%%%%%%%%%%%%%%%%%%%%%%%%%%%%%%%%%%%%%%%%%%%%%%%%

\appendix

\section{Numerical check of the conjectured delay function}

In this appendix, we check that the conjecture in
Eq.~(\ref{eq:conjecture-delay}) is consistent with numerical
calculations.
Our goal is not to present QCD calculations in the kinematics of actual colliders,
but, instead, to check as accurately as possible our theoretical conjecture and
calculations. Therefore, we pick a simple branching random walk model,
and push $y$ to the largest possible values that
allows a calculation of the delay with reasonable computer resources.

As for the specific model, we consider the
well-tested discretization of the branching Brownian motion
introduced in Ref.~\cite{Brunet_2020} and further studied in Ref.~\cite{Le:2020zpy}.
We refer the reader to the latter articles for a complete
description of the model, and in particular, for the numerical values of the constants
that replace $\gamma_0$ and $\chi''(\gamma_0)$ in Eq.~(\ref{eq:conjecture-delay}).

We solve numerically the equivalent BK equation in the form \BKT (which rules
the evolution of $1-\tilde w$):
For our model, it is a discretization of the FKPP
equation. We start
with a sharp step (tantamount to the McLerran-Venugopalan $T$-matrix element),
and evolve
it to the rapidity $y_0$. Then, either we evolve further to the final rapidity $y$
at which we measure the delay,
or we multiply the front by $1-\lambda$, and evolve this new initial condition for
$\tilde y_0\equiv y-y_0$ more time steps.
We eventually compute the difference in the position
between the two fronts we obtain: The number we get is
the ``delay'' we aim at studying.

In the case of a continuous model such as QCD, we could get
this delay by evaluating the integral
\be
\int_{-\infty}^{+\infty} dx \left[\tilde w_0(y,x;y_0)
  -\tilde w_\lambda(y,x;y_0)\right]\equiv f_{y_0;y}^{\text{num}}(\lambda).
\ee
(Note that it has a $y$-dependence since numerical calculations are necessarily
performed for finite $y$).
The integral in the left-hand side is straightforwardly discretized to be taken over
to the model we have implemented.
We repeat the calculation of the delay
for different values of $\lambda$, $y$ and $y_0<y$.

\begin{figure}[h]
  \begin{center}
    \includegraphics[width=.95\textwidth]{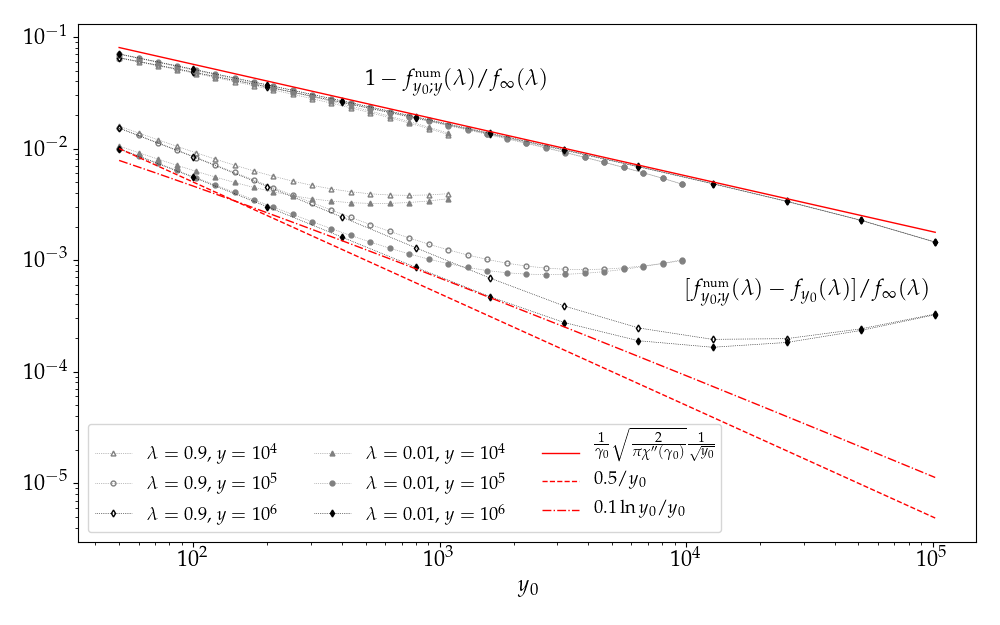}
  \end{center}
  \caption{\label{fig:figure-generating}\small
    Comparison of the
    front delay calculated by solving numerically the exact evolution
    equation and of the conjectured formula~(\ref{eq:conjecture-delay}),
    as a function of $y_0$.
    The points represent the data for
    $1-f^{\text{num}}_{y_0}(\lambda)/f_\infty(\lambda)$, which should tend to
    $\frac{1}{\gamma_0}\sqrt{\frac{2}{\pi\chi''(\gamma_0)}}\frac{1}{\sqrt{y_0}}$
    at large $y$, see Eq.~(\ref{eq:conjecture-delay}) (full line),
    and the distance $[f_{y_0}^\text{num}(\lambda)-f_{y_0}(\lambda)]/f_\infty(\lambda)$
    between the full model and the conjectured asymptotics.
     Two different values of
     $\lambda$ have been considered ($0.9$ and $0.01$), and for each $\lambda$,
     three different values of $y$ ($10^4$, $10^5$, $10^6$).
    The dashed line and dashed-dotted lines are the graphs of
    functions proportional to $1/y_0$ and $\ln y_0/y_0$ respectively.
  }
\end{figure}

The results are shown in Fig.~\ref{fig:figure-generating}.
The upper set of points represents the numerical
data for the delay rescaled by $f_\infty(\lambda)$ and
subtracted from~1 (which is the expected infinite-$y_0$
limit of this rescaled delay). The data is compared to the conjectured
function at finite $y_0$, which reads $\frac{1}{\gamma_0}\sqrt{\frac{2}{\pi\chi''(\gamma_0)}}
\frac{1}{\sqrt{y_0}}$.
We pick two different values of the parameter $\lambda$, namely
$\lambda=0.01$ and $\lambda=0.9$, and for each $\lambda$,
three values of $y$: $10^4$, $10^5$, $10^6$.

We see that in the relevant parametric domain
in which we expect our conjecture to be valid, namely $1\ll y_0\ll y$, all data points
almost superimpose, and approach closely the graph of the conjectured function.
The agreement is better for larger $y$, due to the extension of the
range in $y_0$ of validity of the approximations.

The lower set of points represents the difference of the data and the conjecture.
We see that the mismatch is consistent with a function that decreases with $y_0$
as $1/y_0$, or at most as $\ln y_0/y_0$.\\

Let us mention that we have also checked directly (by solving the equivalent KL equation
for this model)
that the rapidity-gap distribution itself
does indeed converge, for large rapidities, to the predicted form~(\ref{eq:pi}).
We do not report on these numerical calculations here, because
the results look very similar to those for the distribution $G(y,x;y_1)/T(y,x)$
of the splitting rapidity $\tilde y_1$ of the last common ancestor of all dipoles
of log inverse size larger than $x$: The latter
was calculated analytically and numerically in
Ref.~\cite{Le:2020zpy}. The gap distribution and $G/T$
turn out to have the same large-rapidity asymptotics,
except for the extra factor $\ln 2$ present in the expression of the former,
which we have checked numerically to be correct. 
The finite-rapidity corrections exhibit the same patterns.

%%%%%%%%%%%%%%%%%%%%%%%%%%%%%%%%%%%%%%%%%%%%%%%%%%%%%%%%%%%%%%%%%%%%%%%%%%%%%%%%%%%%%%%%%

%\bibliographystyle{h-physrev}
%\bibliography{biblio}

%%%%%%%%%%%%%%%%%%%%%%%%%%%%%%%%%%%%%%%%%%%%%%%%%%%%%%%%%%%%%%%%%%%%%%%%%%%%%%%%%%%%%%%%%

\end{document}